\documentclass[useAMS,usenatbib,twocolumn,floatfix]{mnras}
\usepackage{natbib}
\usepackage[dvipdfmx]{graphicx}	
\usepackage{amsmath}
\usepackage{color}
\usepackage{comment}
\usepackage{multicol}        
\usepackage{pdflscape}	
\usepackage{newtxtext,newtxmath}
\usepackage[T1]{fontenc}
\usepackage{ae,aecompl}

\title{
Insight into primordial magnetic fields from 21-cm line observation
with EDGES experiment
}

\author[T. Minoda et al.]{
Teppei Minoda,$^{1}$\thanks{E-mail: minoda.teppei@d.mbox.nagoya-u.ac.jp}
Hiroyuki Tashiro,$^{1}$
and Tomo Takahashi$^{2}$
\\
$^{1}$Department of Physics and Astrophysics, Nagoya University, Nagoya 464-8602, Japan\\
$^{2}$Department of Physics, Saga University, Saga 840-8502, Japan\\
}

\date{Accepted XXX. Received YYY; in original form ZZZ}

\pubyear{2018}

\begin{document}
\label{firstpage}
\pagerange{\pageref{firstpage}--\pageref{lastpage}}
\maketitle

\begin{abstract}
The recent observation of the 21-cm global absorption signal by EDGES suggests that
the intergalactic medium~(IGM) gas has been cooler than
the cosmic microwave background during $15 \lesssim z \lesssim 20$.
This result can provide a strong constraint on heating sources for the IGM gas at these redshifts.
In this paper we study the constraint on the primordial magnetic fields~(PMFs) by the EDGES result.
The PMFs can heat the IGM gas through their energy dissipation due to the magnetohydrodynamic effects.
By numerically solving the thermal evolution of the IGM gas with the PMFs,
we find that the EDGES result gives a stringent limit on the PMFs as $B_{1\mathrm{Mpc}} \lesssim 10^{-10}$ G.
\end{abstract}



\section{Introduction} \label{sec:intro}
In the Universe, there exist magnetic fields on a wide range of length scales $L$,
from the planets~\citep{Connerney1993} and stars~\citep{Donati2009}
with $L \lesssim 1~\mathrm{au}$,
to the galaxy clusters~\citep{Carilli2001} and large-scale structure~\citep{Vacca2018}
with $L \gtrsim 1~\mathrm{Mpc}$.
Various observations have been revealing the nature of these magnetic fields
(see also~\citealt{2017ARA&A..55..111H, 2018ApJS..234...11H}).
However, the origin and evolution of large-scale magnetic fields are not understood yet
and they are one of the challenging problems in modern cosmology~\citep{Durrer2013}.
Although the dynamo process might have played an important role
in the amplification of large-scale magnetic fields~\citep{Brandenburg2004},
this mechanism cannot create magnetic fields from nothing.
Therefore, many works study the possibility that 
the origin of these magnetic fields could be tiny magnetic fields
created in some physical phenomena in the early universe,
called the `primordial magnetic fields' (PMFs).
In the literature, the generation of the PMFs has been suggested
in various cosmological epochs including
the inflationary era~\citep{Turner1988, Ratra1992},
pre-heating~\citep{Bassett2000},
phase transition~\citep{Hogan1983,Quashnock1989,Baym1995,2017JCAP...07..051A},
topological defects~\citep{Avelino1995,Sicotte1997},
the Harrison mechanism~\citep{Harrison1970,Hutschenreuter2018},
and so on (see, e.g.,~\citealt{Kandus2010,Subramanian2015} for reviews).
Additionally, the possibility of the existence of the intergalactic magnetic fields with $10^{-20}$--$10^{-15} ~ \mathrm{G}$
has been argued by some previous works in this decade~\citep{Ando2010,Neronov2010,Takahashi2013,Tashiro2013},
which motivates the study of the PMFs
as the origin of magnetic fields in galaxy clusters and large-scale structure.

Constraints on the PMFs can provide
the information on their generation mechanism
and, moreover, give a hint to the physics of the early universe.
So far, many constraints on the PMFs
have been obtained from different cosmological observations: 
big bang nucleosynthesis (BBN)~\citep{Cheng1996},
cosmic microwave background (CMB) temperature anisotropy~\citep{PlanckCollaboration2015b}
and its polarisation~\citep{Zucca2016},
the CMB spectral distortion~\citep{Jedamzik1999,Kunze2013},
the baryon-to-photon number constraints
between the BBN and the recombination epoch~\citep{Saga2017},
the Sunyaev-Zel'dovich effect~\citep{Shaw2010,Tashiro2011,Minoda2017},
the galaxy number count~\citep{Tashiro2010},
the star formation history~\citep{Marinacci2015}, and so on.

Now the measurement of cosmological 21-cm line signatures is also
expected to be a useful tool to constrain the PMFs.
The 21-cm line signal depends on the physical states of neutral hydrogen gas
such as its number density, its temperature and so on.
The PMFs can provide the effect on them
through the magnetohydrodynamic (MHD) effects~\citep{Sethi2005}.
\citet{Tashiro2006} have pointed out that
the measurements of 21-cm fluctuations by future radio interferometer telescopes
can provide a strong constraint on the PMFs
and have stimulated further detailed studies~\citep{Shiraishi2014,2019JCAP...01..033K}.

\citet{Bowman2018a} have recently reported
the detection of the strong radio absorption signal around 78 MHz
with the Experiment to Detect the Global Epoch of Reionization Signature (EDGES).
This result indicates that the gas temperature in the intergalactic medium~(IGM)
was cooler than the CMB temperature at the corresponding redshifts,
~i.e., $15 \lesssim z \lesssim 20$.
Therefore, the EDGES result can be interpreted
as a constraint on a cosmological heating source.
Note that the EDGES result is difficult to explain by the standard scenario:
The amplitude of the absorption trough is almost twice as large as
the maximal value expected in the standard cosmology.
However, instead of giving an explanation on this anomaly,
several authors have already applied the EDGES result to constrain some models:
the Hawking evaporation of small primordial black holes (PBHs)~\citep{Clark2018},
the emission from the accretion discs around large PBHs~\citep{Hektor2018a},
the decaying~\citep{Clark2018,Mitridate2018}
or annihilating dark matter~\citep{Cheng2018,DAmico2018}, 
warm dark matter~\citep{Safarzadeh2018}, 
primordial power spectrum~\citep{Yoshiura2018} and so on.

In this work, we study the implication of the EDGES result to the PMFs
and derive a constraint on them.
The PMFs work as extra cosmological heating sources
through the so-called ambipolar diffusion, in particular, in the late universe.
Several authors have already pointed out that the measurement
of the global 21-cm signal can put the constraint on the PMFs
\citep{2009ApJ...692..236S,2009JCAP...11..021S}.
Following these works, we evaluate the global thermal history with the PMFs
and obtain a constraint on the amplitude and scale dependence of the PMFs.

This paper is organized as follows.
In Section~\ref{sec:theory}, we describe the time evolution of the 21-cm line signal,
and we discuss the impact of the PMFs on it in Section~\ref{sec:PMF}.
We report a constraint on the PMFs
from the result of EDGES experiment in Section~\ref{sec:result},
and finally summarize our findings and discuss the future outlook in Section~\ref{sec:conclusion}.
In the analysis of this paper,
we assume a flat lambda cold dark matter ($\Lambda$CDM) model
and fix cosmological parameters 
as obtained by Planck 2015~\citep{PlanckCollaboration2015c}:
$\Omega_\mathrm{m}=0.308, \Omega_\mathrm{b}=0.048,
h \equiv {H_0}/{100~\mathrm{km}~\mathrm{s}^{-1}~\mathrm{Mpc}^{-1}} = 0.678$
where $\Omega_\mathrm{m}$ and $\Omega_\mathrm{b}$
are the density parameters for the matter and baryon
and $h$ is the reduced Hubble constant.

\section{Global 21-cm line signal} \label{sec:theory}
In cosmological 21-cm line measurements,
we observe the differential brightness temperature~\citep{Furlanetto2006},
\begin{eqnarray}
&\delta T_\mathrm{b} (z)
\simeq  27 x_\mathrm{HI}(z) 
\left[1 - \cfrac{T_\gamma(z)}{T_\mathrm{spin}(z)}\right]
\hspace{10mm}
& \nonumber \\
&\hspace{10mm}
\times \left(\cfrac{\Omega_\mathrm{b} h^2}{0.02}\right)
\left(\cfrac{0.15}{\Omega_\mathrm{m} h^2}\right)^{1/2}
\left(\cfrac{1+z}{10}\right)^{1/2}& [\mathrm{mK}]~,
\label{dTb}
\end{eqnarray}
where $x_\mathrm{HI}$ is the neutral fraction of hydrogen,
$T_\gamma$ is the CMB temperature
and $T_\mathrm{spin}$ is the spin temperature
defined by the population ratio of the hyperfine levels in a neutral hydrogen.
Since we are interested in the global 21-cm line signal,
all of these quantities are treated as background ones.

When the spin temperature is the same as the CMB one,
the 21-cm signal vanishes as shown in equation~\eqref{dTb}. 
In the cosmological context,
there are two processes to make the spin temperature deviate from the CMB one.
One is the collisional interaction
and the other is the interaction with Ly $\alpha$ flux field~\citep{Wouthuysen1952,Field1959}.
They can couple the hyperfine structure with the IGM gas temperature.
Besides the CMB interaction~(the 21-cm photon emission and absorption),
these processes control the evolution of the spin temperature.
The spin temperature evolves between the values of the CMB and the IGM gas temperatures.

In the standard cosmology,
the spin temperature evolution can be divided into four regimes.
First, after the decoupling of the gas temperature from the CMB one around $z \sim 200$,
the spin temperature follows the gas temperature through the collisional coupling.
Then at the second stage,
the spin temperature approaches to the CMB one
because the collisional interaction becomes weak due to the cosmic expansion.
In the third regime, the first luminous objects play important roles.
They can produce a strong Ly $\alpha$ field and Ly $\alpha$ interaction becomes effective.
So far the gas temperature evolves adiabatically after the decoupling from the CMB one
and the gas is cooler than the CMB.
Accordingly the spin temperature also gets lower than the CMB one.
Equation~\eqref{dTb} tells us that the global 21-cm signal becomes negative in this regime.
In other words, the signals are observed as the absorption trough.
After that, as the star and galaxy formation becomes active,
a lot of ultraviolet (UV) photons are produced in this regime.
They start to ionize and heat up the gas.
Quickly the gas temperature increases and surpasses the CMB temperature.
The spin temperature also becomes higher than the CMB
and the signal is measured as emission.
Gradually the IGM gas is ionized and, finally, the ionization of the IGM is completed.
At this point the global 21-cm signal totally vanishes again.

Recently, the EDGES experiment has reported the absorption signal of
global redshifted 21-cm lines between redshifts~$15 \lesssim z \lesssim 20$.
This result would indicate that the neutral hydrogen hyperfine structure
coupled well with the gas temperature through Ly $\alpha$ field
and the gas temperature was lower than the CMB one around these redshifts.
In the next section, we show how the PMFs affect the evolution of the IGM gas temperature.

\section{The Formalism of the Primordial Magnetic Fields} \label{sec:PMF}
Since our final aim is to constrain the strength and scale dependence of the PMFs,
we provide their statistical properties.
First, we assume the PMFs as statistically homogeneous and isotropic random Gaussian fields.
The statistical property of such PMFs is completely determined
by only the power spectrum, $P_B(k)$, as~\cite{Landau1980}
\begin{equation}
\langle B^*_i (\mathbf{k}) B_j (\mathbf{k}') \rangle
= \cfrac{(2\pi)^3}{2} \delta_\mathrm{D} (\mathbf{k}-\mathbf{k}')
(\delta_{ij}-\hat{k}_i \hat{k}_j) P_B(k)~,
\end{equation}
where $B^*_i (\mathbf{k})$ is a Fourier component of
$B_i({\mathbf x}) $ with a mode $\mathbf k$ 
and we assume that the PMFs are non-helical fields.

For simplicity,
we assume that the power spectrum of the PMFs
is given by the single power law in $k$ space as
\begin{equation}
P_B(k) = \cfrac{(2\pi)^{n_B+5}}{\Gamma(\frac{n_B+3}{2})} ~ B_n^2~\cfrac{k^{n_B}}{k_n^{n_B+3}}
\hspace{5mm} (\text{for} \hspace{3mm} k<k_\mathrm{cut})~.
\label{P_B}
\end{equation}
Here, we set $k_n \equiv 2\pi$ Mpc$^{-1}$ as a reference scale,
$B_n$ is the amplitude of the PMFs normalized at the length of 1 Mpc,
and $n_B$ gives the scale dependence.
The case of $n_B=-3.0$ corresponds to the scale-invariant PMFs
and we adopt $n_B>-3.0$ throughout this paper to avoid the infrared divergence of the PMFs
and $\Gamma(x)$ represents the gamma function.

On the other hand, in the UV regime,
the PMFs have the cut-off scale due to the MHD effect in the early universe.
According to previous studies~\citep{Jedamzik1996,Subramanian1997},
the radiative diffusion before the recombination epoch damps the PMFs on small scales.
The damping scale increases as the universe evolves.
Therefore, we assume that the power spectrum has the sharp cut-off
on the damping scale at the recombination epoch,
\begin{equation}
P_B(k)=0 \hspace{10mm} (\text{for} \hspace{3mm} k \ge k_\mathrm{cut})~.
\end{equation}

We consider the time evolution of the cut-off scale
as $k_\mathrm{cut}(t) = k_\mathrm{cut, init} f(t)$ with $f(t_\mathrm{init})=1$.
Here $k_\mathrm{cut, init}$ is the cut-off scale
at the initial time (the recombination epoch) as~\citep{2002PhRvD..65l3004M},
\begin{eqnarray}
\left( \cfrac{k_n}{k_\mathrm{cut, init}} \right)^2
=\cfrac{V_\mathrm{A}^2}{\sigma_\mathrm{T}}
\displaystyle \int^{t_\mathrm{rec}}_0 \cfrac{dt}{a^2(t) n_\mathrm{e}(t)}
\hspace{30mm} \nonumber \\
\simeq \left[1.32 \times 10^{-3} \left(\cfrac{B_n}{1~\mathrm{nG}}\right)^2
\left(\cfrac{\Omega_\mathrm{b} h^2}{0.02}\right)^{-1}
\left(\cfrac{\Omega_\mathrm{m} h^2}{0.15}\right)^{1/2}
\right]^{2\over{(n_B+5)}}~,
\label{cut}
\end{eqnarray}
where $V_\mathrm{A}$, $\sigma_\mathrm{T}$, $n_\mathrm{e}$ and $t_\mathrm{rec}$
are the Alfv\'en velocity, the cross-section for Thomson scattering,
the electron number density, and the cosmic time at the recombination epoch, respectively.
The time evolution of the cut-off scale is represented as $f(t)$.
We will discuss the evolution $f(t)$ later.

In constraining the PMFs, it is useful to introduce $B_\lambda$,
which represents the field strength smoothed at any spatial length $\lambda$.
By choosing the Gaussian window function in Fourier space,
$B_\lambda$ is related to $B_n$ and $k_\lambda=2\pi/\lambda$ as
\footnote{
We have used this definition of the smoothed amplitude of the PMFs
from~\cite{PlanckCollaboration2015b}.
Although this definition is slightly different from some other works
(such as \citealt{Fedeli2012, Saga2017}),
the value of $B_\lambda$ is almost unchanged among different definitions.
}
\begin{eqnarray}
B_\lambda^2 = \int^{\infty}_0 \mathrm{e}^{-k^2 \lambda^2} P_B(k)~ \frac{d^3k}{(2\pi)^3}
= B_n^2 \left(\cfrac{k_\lambda}{k_n}\right)^{n_B+3}~.
\label{eq:Bamp_sm}
\end{eqnarray}

After the recombination epoch, the magnetic fields dissipate their energy
and heat the IGM gas through two processes~\citep{Sethi2005}.
One is the decaying turbulence (DT) and the other is the ambipolar diffusion (AD).
It has been shown that
while the first one is active around the recombination epoch,
the latter becomes effective at the late universe~($z<500$).
The evolution of the IGM gas temperature $T_\mathrm{K}$
with the heating from the PMFs is given by
\begin{equation}
\cfrac{dT_\mathrm{K}}{dt} = -2H T_\mathrm{K}
+ \cfrac{x_\mathrm{e}}{1+x_\mathrm{e}}
\cfrac{8\rho_\gamma \sigma_\mathrm{T}}{3m_\mathrm{e} c} (T_\gamma - T_\mathrm{K})
+ \cfrac{2}{3k_\mathrm{B} n_\mathrm{b}} (\dot{Q}_\mathrm{AD} + \dot{Q}_\mathrm{DT}) ~,
\label{T_gas}
\end{equation}
where $H$, $x_\mathrm{e}$, $\rho_\gamma$, $m_\mathrm{e}$, $c$, $k_\mathrm{B}$ and $n_\mathrm{b}$
is the Hubble parameter, the ionization fraction of the baryon gas, the energy density of CMB photons,
the rest mass of an electron, the speed of light, the Boltzmann constant,
and the number density of the baryon gas, respectively.
$\dot{Q}_\mathrm{AD}$ and $\dot{Q}_\mathrm{DT}$ on the right-hand side represent
the global heating rate due to the ambipolar diffusion and the decaying turbulence, respectively.
These two heating rates are written as~\citep{Sethi2005}
\begin{eqnarray}
\dot{Q}_\mathrm{AD} = \cfrac{|(\nabla \times \mathbf{B}) \times \mathbf{B}|^2}{16\pi^2 \xi \rho_\mathrm{b}^2} \cfrac{1-x_\mathrm{e}}{x_\mathrm{e}}~,
\hspace{33mm}
\label{gamma_pmf_ad} \\
\dot{Q}_\mathrm{DT} = \cfrac{3w_B}{2} H \cfrac{|\mathbf{B}|^2}{8\pi}a^4
\cfrac{\left[\ln(1+t_\mathrm{d}/t_\mathrm{rec})\right]^{w_B}}{\left[\ln(1+t_\mathrm{d}/t_\mathrm{rec}) + \ln(t/t_\mathrm{rec})\right]^{1+w_B}}~,
\label{gamma_pmf}
\end{eqnarray}
with the time dependence of the decaying turbulence $w_B \equiv {2(n_B + 3)}/{(n_B + 5)}$,
the mass density of the baryon gas $\rho_\mathrm{b}$,
the physical time-scale for the decaying turbulence $t_\mathrm{d}$,
and the drag coefficient
$\xi=1.9~\left({T_\mathrm{K}}/{1~\mathrm{K}}\right)^{0.375}
\times10^{14}~\text{cm}^3 \text{g}^{-1} \text{s}^{-1}$
as referred to in~\citet{2008PhRvD..78h3005S}.
Here, since we assume a blue spectrum of the PMFs, $n_{\rm B}>-3$,
we can take $t_\mathrm{d} = (k_\mathrm{cut}V_\mathrm{A})^{-1}$
as the Alfv\'en time-scale at the cut-off scale of the PMFs, as referred to in~\citet{Sethi2005}.
Also, we can calculate the absolute value of the Lorentz force and the magnetic energy
in equations~\eqref{gamma_pmf_ad} and \eqref{gamma_pmf} at a time $t$
by substituting equation~\eqref{P_B} with $f(t)$ into
$|(\nabla \times \mathbf{B})\times\mathbf{B}|^2
= \int (dk_1/2\pi)^3 \int (dk_2/2\pi)^3 k_1^2 P_B(k_1) P_B(k_2)~f^{2n_B+8}(t)(1+z)^{10}$
and $|\mathbf{B}|^2 = \int (dk/2\pi)^3 P_B(k)~f^{n_B+3}(t)(1+z)^4$.
In this study, we neglect following two effects on the thermal evolution of the IGM gas
included in the previous work~\citep{2008PhRvD..78h3005S}.
First, we do not take into account any radiative cooling effects,
such as collisional excitation and ionization, recombination, and bremsstrahlung.
We have confirmed these terms are negligible for the PMF model parameters of our interest.
The other assumption is that there are no astrophysical objects.
We mention this point in Section~\ref{sec:conclusion}.

In order to solve equation~\eqref{T_gas},
we also need to follow the evolution of the ionization fraction,
\begin{eqnarray}
\cfrac{dx_\mathrm{e}}{dt} = \gamma_\mathrm{e} n_\mathrm{b} x_\mathrm{e}
+ \cfrac{1 + K_\alpha \Lambda n_\mathrm{b}(1-x_\mathrm{e})}
{1 + K_\alpha (\Lambda+\beta_\mathrm{e}) n_\mathrm{b}(1-x_\mathrm{e})}
\hspace{15mm}
\nonumber \\
\times \left[-\alpha_\mathrm{e} n_\mathrm{b} x_\mathrm{e}^2 + \beta_\mathrm{e} (1-x_\mathrm{e})
\exp \left(-\cfrac{3 E_\mathrm{ion}}{4 k_\mathrm{B} T_\gamma}\right)\right]~,
 \label{x_e}
\end{eqnarray}
where $K_\alpha$, $\Lambda$, $\alpha_\mathrm{e}$, $\beta_\mathrm{e}$
and $\gamma_\mathrm{e}$ are the parameters for the ionization and the recombination processes.
For these parameters,
we adopt the functions in~\citet{Seager1999} and~\citet{Seager2000}
with the modifications suggested in~\citet{Chluba2015}.
In equations~\eqref{T_gas} and~\eqref{x_e},
we neglect the primordial helium and the heavier elements for simplicity.

Finally, we comment on the time evolution of
the cut-off scale $f(t)$.
Since the dissipation of the PMFs is effective on smaller scales,
the dissipation could evolve the cut-off scales.
In this study, we obtain the evolution of $f(t)$
based on the energy conservation of PMFs
including the energy dissipation
given in equations~\eqref{gamma_pmf_ad} and \eqref{gamma_pmf},
\begin{eqnarray}
\cfrac{d}{dt} \left(\cfrac{|\mathbf{B}|^2}{8\pi}\right)
= -4H\cfrac{|\mathbf{B}|^2}{8\pi} - \dot{Q}_\mathrm{AD} - \dot{Q}_\mathrm{DT}~.
\label{E_mag}
\end{eqnarray}
Rewriting this equations in terms of the PMF power spectrum
with the cut-off scales, we can obtain the time evolution of $f(t)$.

Now, assuming the power spectrum of the PMFs,
we solve equations~\eqref{T_gas},~\eqref{x_e} and~\eqref{E_mag} simultaneously,
and we can uniquely obtain the evolution of $T_\mathrm{K}$, $x_\mathrm{e}$, and $f(t)$
for a given PMF model with $(n_B, B_n)$.

\section{Results} \label{sec:result}
\begin{figure}
\centering
\includegraphics[width=\columnwidth]{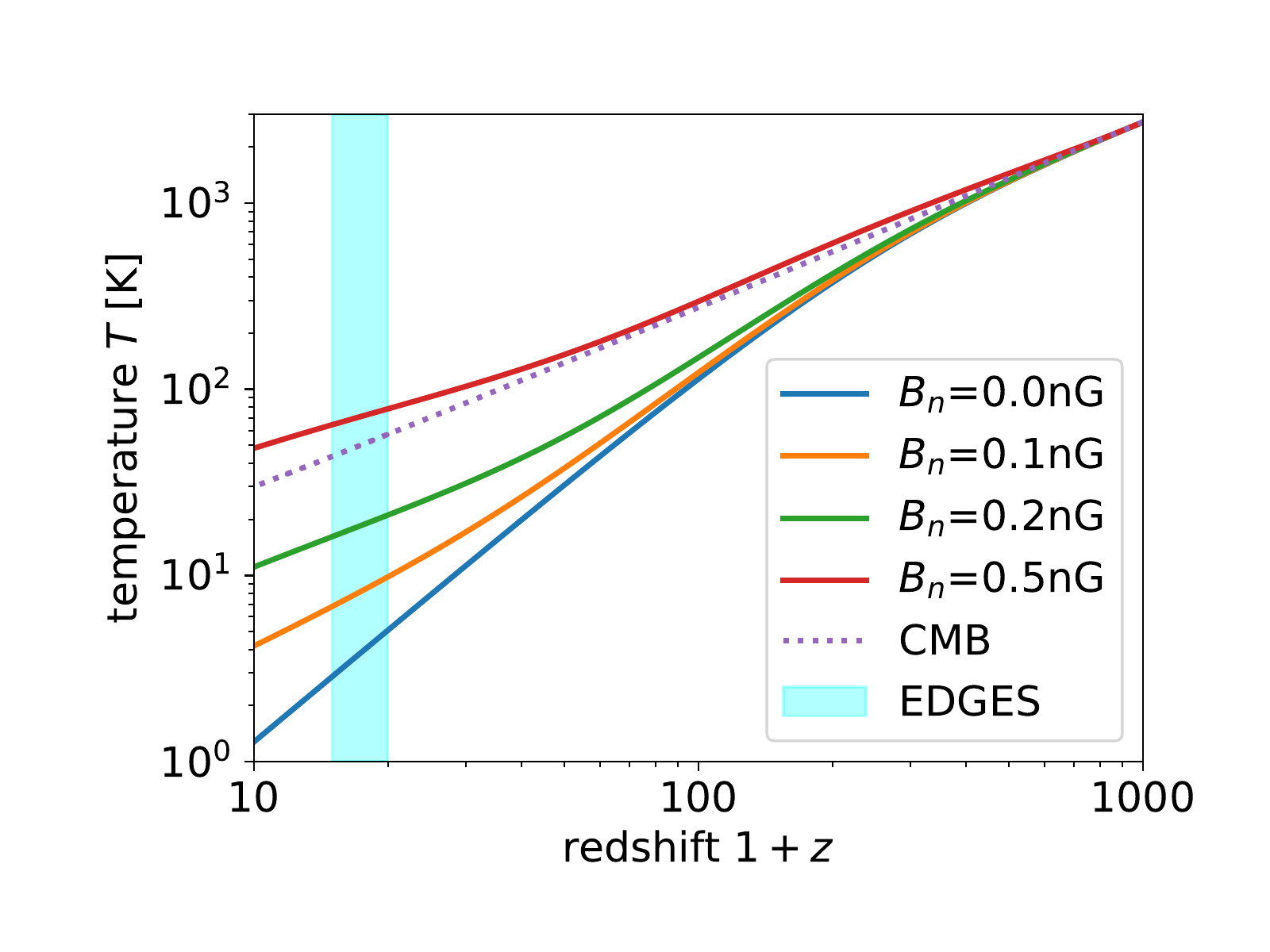}
\vspace{-10mm}
\caption{
The gas temperature evolutions with the PMFs.
The dotted line is the CMB temperature,
and solid lines are the gas temperature for the cases with $B_n=0.0, 0.1, 0.2, 0.5$ nG, respectively.
In this figure, $n_B$ is fixed to $-2.9$ for these lines.
The blue shaded region represents the redshift range $15 \lesssim z \lesssim 20$
corresponding to the strong absorption signal reported by EDGES.
}
\label{T_evo}
\end{figure}

We show the evolution of the gas temperature $T_\mathrm{K}$ for different $B_n$
with the nearly scale-invariant PMFs ($n_B=-2.9$) in Fig.~\ref{T_evo}.
Even with the sub-nano Gauss PMFs,
$T_\mathrm{K}$ is strongly affected by the PMFs.
In this case,
$T_\mathrm{K}$ starts to be heated around $z \sim 200$,
and is well deviated from the one in the case without the PMFs~($B_n =0$ nG).

We have confirmed that the ambipolar diffusion contributes
to the heating of the IGM more than the decaying turbulence at the redshifts of our interest.
Also the redshift dependence of the heating term in equation~\eqref{gamma_pmf_ad} follows as
$2 \dot{Q}_\mathrm{AD}/3k_\mathrm{B} n_\mathrm{b} \propto (1+z)$.
On the other hand,
the dominant cooling term on the right-hand side of equation~\eqref{T_gas} for $z \ll 200$
is the first one,
and $-2HT_\mathrm{K} \propto (1+z)^{3.5}$,
which decreases faster than the heating term due to the ambipolar diffusion.
Therefore, as we take larger values for $B_n$,
the heating due to the ambipolar diffusion gets effective
and dominates the cooling terms at higher redshifts.
After the IGM is heated well and the magnetic energy is considerably dissipated,
the heating source terms in equation~\eqref{gamma_pmf_ad} become decreased.
As a result, the gas temperature starts to deviate from the adiabatic evolution ($B_n=0$ nG) at first,
and the IGM is gradually getting cool after the saturation.
We find the larger value of $B_n$ makes the earlier and stronger heating,
and the value of $n_B$ determines the duration of the PMF heating.

\begin{figure}
\centering
\includegraphics[width=\columnwidth]{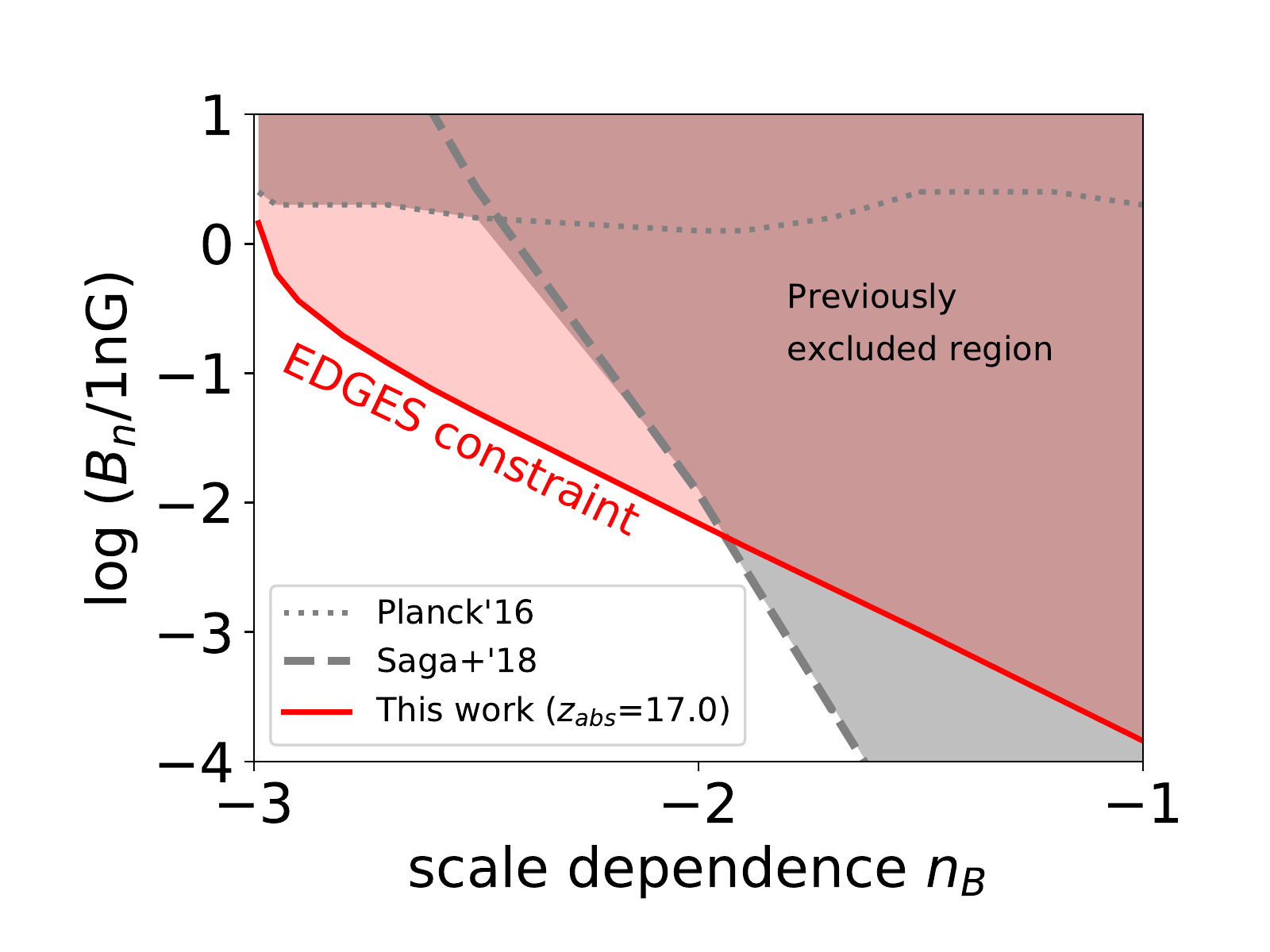}
\vspace{-10mm}
\caption{
The upper limit of the PMFs parameters obtained by
the recent observation of 21-cm line absorption by EDGES experiment (solid).
The gray shaded region is the excluded region by previous works:
\citet{PlanckCollaboration2015b} in the dotted line and \citet{Saga2017} in the dashed line.
}
\label{const}
\end{figure}

The EDGES experiment has reported the detection of the global 21-cm absorption signals
in the redshifts range $15 \lesssim z \lesssim 20$~\citep{Bowman2018a}.
In Fig.~\ref{T_evo} we highlight this redshift region with the blue shade.
To create the absorption signal,
the IGM gas temperature should be lower than the CMB temperature
as shown in equation~\eqref{T_evo}.
Therefore, when the absorption signal is detected at the redshift~$z_{\rm abs}$,
we can exclude a PMF parameter region~$(n_B, B_n)$
which gives $T_\mathrm{K} > T_\gamma$ at $z_{\rm abs}$.
According to the EDGES result,
we set $z_{\rm abs}=17$ which is the central redshift of the absorption signal
in the EDGES experiment.
Calculating the evolution of the gas temperature for different $B_n$ with a fixed $n_B$,
we get a novel constraint on the PMFs from the condition
$T_\mathrm{K} < T_\gamma$ at $z_{\rm abs}=17$.
The obtained constraints are
$B_n \lesssim 1.2 \times 10^{-1}~\mathrm{nG}$ for $n_B=-2.9$,
$B_n \lesssim 6.3 \times 10^{-3}~\mathrm{nG}$ for $n_B=-2.0$, and
$B_n \lesssim 2.0 \times 10^{-4}~\mathrm{nG}$ for $n_B=-1.0$.
We can fit our new constraint in a linear relation between $B_n$ and $n_B$ as
\begin{equation}
\log \left(\frac{B_n}{1~\mathrm{nG}} \right) \lesssim -\frac{(3n_B +10)}{2}
\quad {\rm  for}~ -3.0<n_B<-1.0. 
\end{equation}
We plot our PMF constraint on the $(n_B$, $B_n)$ plane in Fig.~\ref{const} with a solid line.
For comparison, we also show the constraint from~\citet{PlanckCollaboration2015b}
and the one from the magnetic reheating before the recombination~\citep{Saga2017}
by the dotted and dashed lines, respectively.
In the range of $-3.0<n_B<-2.0$,
the EDGES experiment could constrain the PMF amplitude most tightly.

Finally, by varying $z_{\rm abs}$ from 15 to 20,
we have investigated the dependence of the PMF constraint on $z_{\rm abs}$.
We have found out that the resultant upper limit
differs less than 10 per cent for different $z_{\rm abs}$ between 15 and 20.
We can conclude that the dependence on $z_{\rm abs}$ is very weak.

\section{Conclusion} \label{sec:conclusion}
In this paper, we obtained a novel constraint on the PMFs
from the result of the EDGES 21-cm global signal.
The EDGES has reported the detection of the 21-cm absorption signals
in the redshifts between $15 \lesssim z \lesssim 20$.
This result suggests that the IGM gas is cooler than the CMB during this epoch.
The PMFs can heat up the IGM gas by the dissipation of their energy
through ambipolar diffusion and decaying turbulence.
Therefore, the EDGES result can provide the constraint on the PMFs.

We have numerically evaluated the thermal evolution of the IGM gas with the PMFs.
By requiring $T_\mathrm{K} < T_\gamma$ at $z_\mathrm{abs} = 17.0$
at which the EDGES absorption profile is centred,
we have obtained a stringent upper bound on the PMFs,
roughly about $B_n \lesssim 0.1~\mathrm{nG}$.
We also find that this PMF constraint is not changed vey much
if the absorption redshift is deviated from $z_\mathrm{abs} = 17.0$.

To explain the EDGES anomaly,
some non-standard cooling mechanisms are required,
e.g.~\citet{2014PhRvD..90h3522T} and \citet{2018Natur.555...71B}.
If we include such cooling in our analysis,
our constraint on the PMFs should be relaxed.
However, the IGM heating rates (equations \ref{gamma_pmf_ad} and \ref{gamma_pmf})
is proportional to $B_n^4$ and $B_n^2$.
Therefore, our constraint would not be modified very much
even if we considered some non-standard cooling mechanisms.

To end this paper, we briefly discuss the impact of the astrophysical objects on the PMF constraint.
Of course, they can significantly heat up and ionize the IGM gas,
and the constraint from the 21-cm absorption signal might become tighter
if we include such astrophysical processes
(e.g. the star formation and the active galactic nucleus activity).
On the other hand, some previous works have pointed out that
the condition of the astrophysical object formation
is also affected by the PMFs~\citep{Wasserman1978, Shibusawa2014}.
In order to calculate the global 21-cm signal history with these effects,
we should solve the fully non-linear MHD equations
and the realistic astrophysical processes~\citep{Minoda2017,2019JCAP...01..033K}.
We leave these points to future work.

\section*{Acknowledgements}
We thank Kazuhiro Kogai, Hidenobu Yajima and Kenji Hasegawa for useful comments.
This work is supported by KAKEN
No. 15K17646~(HT), 17H01110~(HT) 15K05084~(TT), 17H01131~(TT)
and MEXT KAKENHI Grant Number 15H05888~(TT).

\bibliographystyle{mnras}

\bibliography{library}

\bsp	
\label{lastpage}
\end{document}